# Young Stars Discovered in Dwarf Spheroidal Galaxies Confirm their Recent Infall into the Milky Way

François Hammer[1]
Piercarlo Bonifacio[1]
Elisabetta Caffau[1]
Yanbin Yang[1]
Frédéric Arenou[1]
Carine Babusiaux[2]
Monique Spite[1]
Patrick François[1]
Ana Gomez[1]
David Katz[1]
Lorenzo Monaco[3]
Marcel Pawlowski[4]
Jianling Wang[1]

[1] LIRA, Paris Observatory – PSL, CNRS, France
[2] Grenoble Observatory, CNRS, France
[3] Andres Bello University, Concepcòn, Chile
[4] Leibniz Institute for Astrophysics, AIP, Potsdam, Germany

Recent observations from ESA's Gaia satellite and with ESO's Very Large Telescope have identified the presence of a population of young stars, 0.5 to 2 Gyr old, in the halo of, and in dwarf spheroidal galaxies surrounding, the Milky Way (MW). It suggests that MW dwarf galaxies, currently devoid of gas, had, until recent times, enough gas to sustain a burst of star formation. The recent loss of gas coincides with their arrival in the vicinity of the MW, in agreement with orbital predictions from Gaia that indicate that most dwarf galaxies reached the MW halo less than 3 Gyr years ago. This completely changes the interpretation of their dynamics, mass and dark matter content.

## A recent infall of most dwarf galaxies is predicted by the hierarchical scenario

In the last 30 years, astronomers have conducted extensive observations and analyses of the stellar populations in the dwarf galaxies surrounding the Milky Way (MW) galaxy. Several dwarf galaxies were thought to be made up of only very old stars (with ages much greater than 6 Gyr) with a low concentration of elements heavier than helium. It was then deduced that these dwarf galaxies, such as the Sculptor dwarf spheroidal (dSph), had lost their gas at these remote epochs, when they became satellites of our Galaxy, orbiting around it ever since. This scenario has a major consequence in near-field cosmology: these dwarf galaxies must have a huge quantity of dark matter in order to protect their stellar content from the disruptive force of the MW's gravitational field. Indeed, in the absence of a large amount of dark matter, tidal forces from the MW would disperse the stars of the dwarf galaxy in just a few Gyr. Until now, they have been considered as the most dark-matter-dominated galaxies in the Universe, whose total masses, derived from their large velocity dispersions, are 10 to 1000 times larger than their stellar masses.

Gaia observations (the second and third data releases, DR2 & DR3) provided detailed orbital motions for 156 globular clusters (GCs; Vasiliev, 2019), and for 46 MW dwarf galaxies (Li et al., 2021), allowing their orbital (or binding) energies to be accurately calculated. Several studies (Kruijssen et al., 2019, 2020; Massari, Koppelman & Helmi, 2019; Malhan et al., 2022) showed that several GCs are associated with the important accretion events that occurred in the MW, namely the elaboration of the bulge (12–13 Gyr ago), the Kraken (11–12 Gyr ago) and the Gaia-Sausage-Enceladus (GSE, 8–10 Gyr ago) merger events, and more recently the dwarf Sagittarius galaxy (Sgr, 4–6 Gyr ago) infall into the Galactic halo. These associations have been determined after comparing GC ages, metallicities and the orientation of their orbital angular momenta and binding energy to those of stars from which the above events have been identified. They are expected since stars and GCs are often formed together through gas-rich major-merger events (De Lucia et al., 2024; Valenzuela et al., 2024).

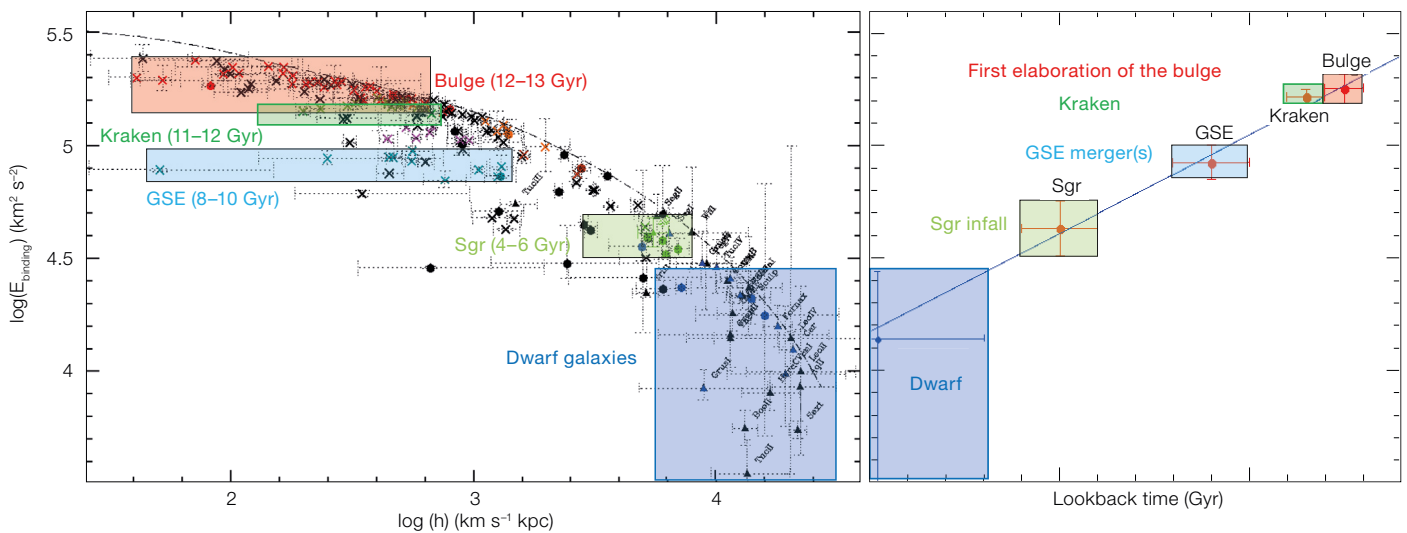

Figure 1. The left panel shows total energy versus angular momentum ($h = R_{GC} \times V_{tan}$) on a logarithmic scale, for high-surface-brightness GC (crosses), low-surface-brightness GC (filled circles), and dwarf galaxies (triangles). Structures identified by Malhan et al. (2022) and Kruijssen et al. (2020) are added in different colours. VPOS (Vast POlar Structure; see Pawlowski, Pflamm-Altenburg & Kroupa, 2012 and Li et al., 2021) dwarf galaxies are shown in blue. The dot-dashed line shows the limit that cannot be passed by any orbits, as it is fixed for a circular orbit (the largest possible binding energy for a given angular momentum). The right panel shows the corresponding lookback time of stellar system entry in the Milky Way halo as a function of its current binding energy for different families of GCs, and for the dwarf galaxies that do not belong to the tightly bound Sgr system (that is, excluding Sgr, Segue I, II, Tucana III, IV and Willman I). The blue solid line is a linear fit. A simple interpretation is that dwarf satellites with $\log(E_{binding}/(\text{km}^2\ \text{s}^{-2})) < 4.34$ are on their initial approach (see blue box), a value close to the logarithm of the average energy (4.14 km² s⁻²) of dwarf galaxies, whose scatter provides an upper limit of $E_{binding} = 4.34$. The latter combined with the linear fit suggests a lookback time of halo entry of dwarf galaxies less than 3 Gyr.

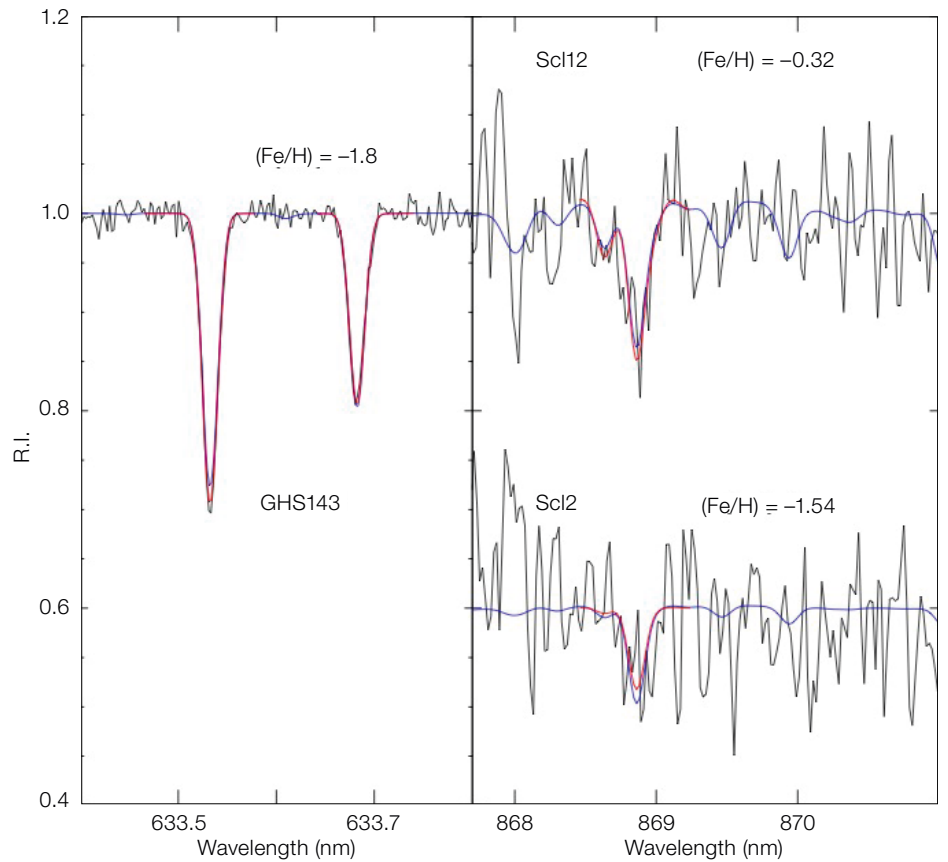

Figure 2. Left: UVES spectrum of the young metal-poor star GHS143, for which the two Fe I lines indicate a low-metallicity star. The blue line is the synthetic spectrum corresponding to the average metallicity and the red line is the best fit to each line. The high proper motions imply that the star is unbound and falling into the Galaxy, its estimated age being between 5 and 9 Myr. Right: GIRAFFE spectra of two young stars in Sculptor, for which the 868.86 nm Fe I line indicates a difference of over 1 dex in iron abundance. The spectrum of Scl12 has been arbitrarily shifted by –0.4, for display purposes. The blue line is the synthetic spectrum corresponding to the average metallicity, while the red line is the best fit to the line.

Figure 1 shows the distribution of GCs (crosses for the high-surface-brightness examples, large dots for the low-surface-brightness ones; see Hammer et al., 2021) and of dwarf galaxies (triangles) in the binding energy–angular momentum plane. GCs are coloured on the basis of the structures to which they belong. Each structure (bulge, Kraken, GSE, and Sgr) shows a very narrow range in binding energy, which allows them to be identified in the relation between binding energy and infall lookback time.

Galaxies like the MW follow a hierarchical structure formation, in which smaller galaxies merge into larger systems over time. This 'inside-out' growth pattern means that older structures are tightly bound to the galaxy, while more recent arrivals are less so. Hammer et al. (2023) derived the MW's accretion history (see Figure 1), which can be fitted by a single line from bulge elaboration to Sgr. The line slope is in agreement with predictions from high-resolution cosmological simulations (Rocha, Peter & Bullock, 2012; and see a detailed comparison to simulations in Hammer et al., 2024a). By extrapolating the line to the lower binding energy of dwarf galaxies, it suggests that most of the latter galaxies have reached the MW halo less than 3 Gyr ago.

### Discovery of young stars in the halo and then in dSph galaxies

Thanks to the proper motions provided by Gaia, Bonifacio et al. (2024) and Caffau et al. (2024a,b) identified several stars in the MW halo with high velocities with respect to the Sun (> 500 km s$^{-1}$). 10 of these high-velocity stars show large masses (in excess of 1.3 $M_\odot$, half of them in excess of 2 $M_\odot$), young ages (0.3 to 2.5 Gyr old) and metallicities from –1.3 to –2.2. Figure 2 shows the Ultraviolet and Visual Echelle Spectrograph (UVES) spectrum of GHS 143 with a mass of 2.3 $M_\odot$. Measurements of metallicity from spectroscopy combined with accurate Gaia photometry allow an accurate determination of the star's mass and age[a]. It is well known that some stars may 'rejuvenate' by mass transfer from a binary companion or even by merging with another star; such a rejuvenated star is called a blue straggler, because of its anomalous position in the colour–magnitude diagram. However, the formation of blue stragglers with masses larger than two solar masses is unlikely because most halo stars are > 8 Gyr old, for which the maximum mass at the turn-off is expected to be 0.9 $M_\odot$. To explain young and massive halo stars through the blue straggler channel would require mergers involving three stars, which are extremely rare events. The simplest explanation of the data is that these stars are not rejuvenated, but are genuinely young.

The origin of these halo young stars is a puzzle, since the halo does not contain large amounts of gas to fuel star formation. It is thus likely that they are formed during the earlier infall of a dwarf galaxy, for which the original gas content has been ram-pressurised during its interaction with the Galactic corona. It has prompted us to verify whether or not there might be also a population of young stars in the most massive dwarfs, i.e., dSph galaxies. Yang et al. (2024) have been able to 'filter' the dSph members by excluding the foreground stars (from the MW) that appear in the field of view of the dSph galaxies. This filtering is extremely efficient, and was not possible before the availability of the Gaia data. For example, in the field of view of Sculptor, half of the stars are foreground, but thanks to the filtering using the Gaia proper motions, this contamination is at most 1.4%! This filtering method allowed the unambiguous identification of young stars in Sculptor whose ages are between 0.5 and 2 Gyr and whose masses are up to three times that of the Sun, after determining their metallicities through observations with the GIRAFFE spectrograph (see Figure 2). Besides confirming that these stars belong to Sculptor, spectroscopic observations allow their masses to be derived. Because the turnoff (TO) point of the Sculptor dSph is similar to that of the halo (0.9 $M_\odot$), it is also unlikely that Sculptor young stars can be blue stragglers. Besides Sculptor, three additional dSphs, Sextans, Ursa Minor and Draco, show the presence of a young star population, demonstrating that the phenomenon is fairly common, especially because other dSphs (Fornax, Carina, Leo I and Leo II) are also known to contain young populations. The most favourable location in the colour-magnitude diagram to identify young stars is the so-called 'yellow plume' (see, for example, Gullieuszik et al., 2008), an almost vertical sequence to the blue of the red giant branch. These stars are three to four magnitudes brighter than the TO and are young evolved stars in the core helium-burning phase, or young sub-giant stars. The same region can also be occupied by evolved blue stragglers, but such an interpretation becomes unlikely for stars of mass above two solar masses.

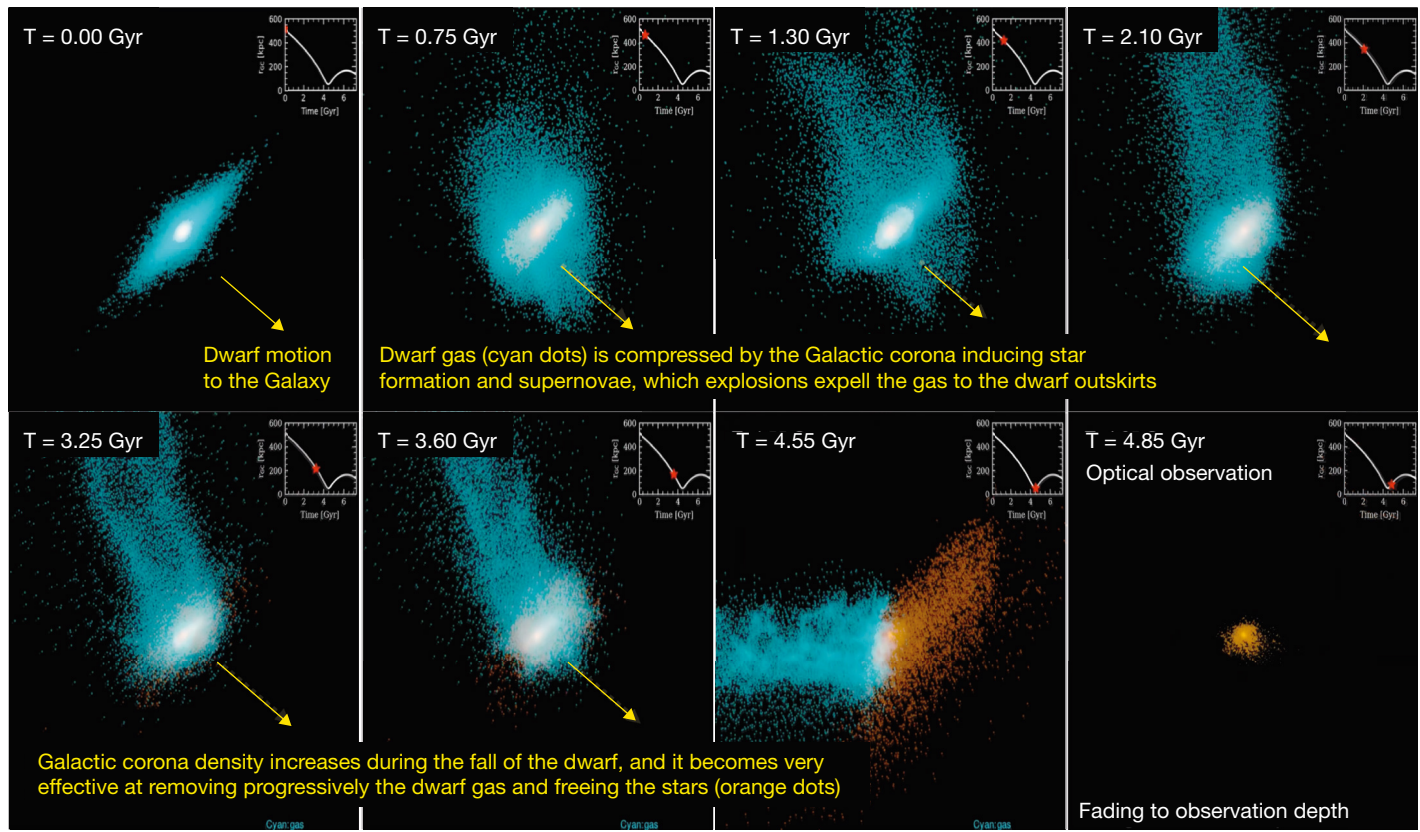

Figure 3. Snapshots of a video summarising the hydrodynamical simulation of the infall of a gas-rich dwarf galaxy that is ram-pressurised during its infall into the MW corona (Wang et al., 2024). Gas is represented by cyan particles and stars by red particles. The yellow arrow indicates the dwarf motion, and the insert on the top-left the radial evolution with time. The final galaxy in the bottom-right panel has properties very similar to the Sculptor dSph.

Conclusions

The discovery of young stars in four dSph galaxies formerly considered as uniquely made of old stars changes our view of the history of their motions relative to the MW. It supports the scenario of a recent infall suggested by the precise determination of their orbits from Gaia. In fact, since stars are known to form within gas clouds, the dwarf galaxies must have contained sizable amounts of gas up to 0.5 to 2 Gyr years ago. It is well accepted that MW dwarf progenitors were gas-rich galaxies similar to present-day dwarf irregular galaxies observed in the field. Once they arrived in the MW halo, their gas was stripped by the ram pressure exerted by the hot gas in the MW corona. The process is quite rapid, because masses of dwarf galaxies are three to five orders of magnitude smaller than that of our galaxy.

The above discoveries have considerable consequences for our understanding of the nature of dwarfs surrounding the MW. Hydrodynamical simulations (Wang et al., 2024) show that during the interaction with the MW corona gas, the stellar content is considerably shaken by turbulence effects during the process (see Figure 3[b]). dSph progenitors gradually lost their gas, provoking a strong disequilibrium in the residual stellar content, including by tidal shocks exerted by the MW's gravity (Hammer et al., 2024b). When the gas is fully lost, stars begin to expand thanks to the associated loss of internal gravity, and this naturally explains the presence of stars associated with many dwarfs while being very far from their centres (Chiti et al., 2023; Sestito et al., 2023; Longeard et al., 2022; Waller et al., 2023). Hydrodynamical simulations predict that, in the case of a recent arrival, a small number of stars must be formed during the interaction between the gas-rich progenitor and the MW corona, which is consistent with the observed number of young stars (Yang et al., 2024). They also reproduce all the properties of dwarf galaxies, including their observed velocity dispersions, with a very limited amount of dark matter, or even without dark matter at all.

Our results are suggestive and require further confirmation. In particular spectroscopic observations at higher signal-to-noise ratio of more yellow plume stars in Sculptor and other dwarf galaxies are needed. Young stars with masses in excess of two solar masses have rotational velocities in excess of 200 km s$^{-1}$ when on the main sequence. As they evolve they slow down, but they can still show measurable rotational velocities of the order of 10–20 km s$^{-1}$ (Lombardo et al. 2021). Evolved blue stragglers, instead, can spin up to 100 km s$^{-1}$ at the time of mass transfer, but spin down rapidly, even more so as they evolve. At the same time we need to obtain high-resolution, high-signal-to-noise spectra for the many halo yellow-plume stars, that we can select from Gaia. Their metallicity and dynamics should trace their origin to an existing or disrupted dwarf galaxy.

If confirmed, this novel paradigm for explaining the observations of MW dwarf galaxies will become a serious contender to the scenario of dark matter-dominated MW dwarf galaxies, and this could lead us into a new area in near-field cosmology.

### Notes

[a] Determination of metallicity through spectroscopy allows the appropriate isochrones in the colour–magnitude diagram to be chosen, and hence the stars' ages and masses to be estimated.
[b] A video of the simulation of the Sculptor dSph (Wang et al., 2024) shown in Figure 3 can be seen at https://www.youtube.com/watch?v=SwxSdmfQis4

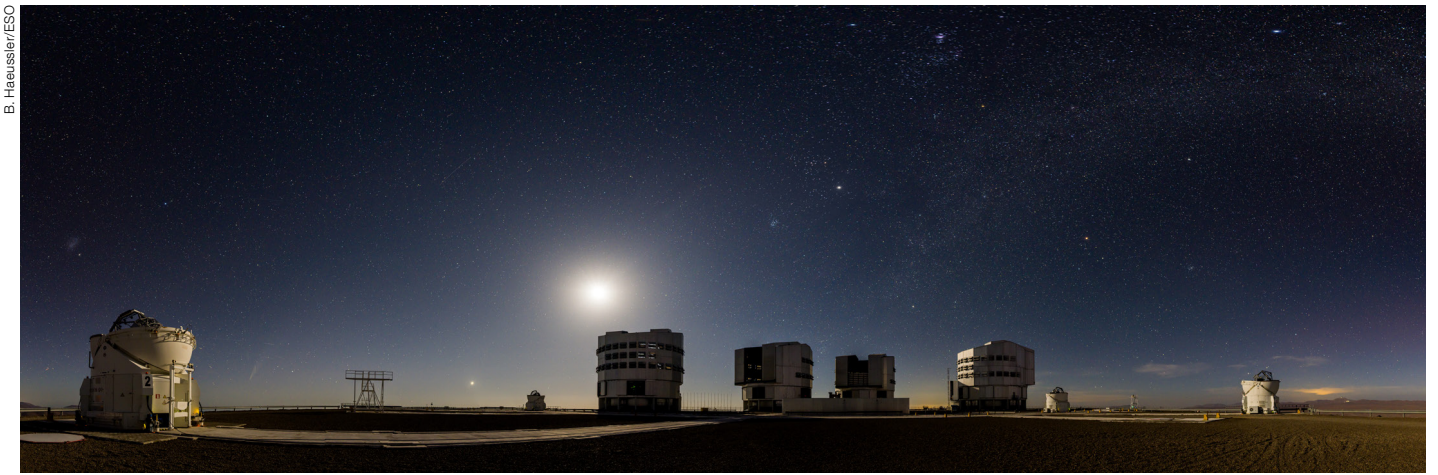

B. Haeussler/ESO

In this portrait of ESO's Paranal Observatory, taken in early February, our planets appear to parade one after the other across the night sky. In addition to the Moon, our own Milky Way, and the comet C/2024 G3, we can see Saturn, Venus, Jupiter and Mars — even Neptune and Uranus are hiding here too! Often on nights with a few planets in view, you can draw an imaginary straight line in the night sky through them. This is due to their orbital paths being relatively aligned along a single, flat plane called the ecliptic. (In reality, the planets aren't aligned one after the other in a straight line in the Solar System, they are fanned out; but we can still see them simultaneously in the sky, which only happens every few years.) You may notice that in this image the planets are not contained within the band of the Milky Way, and that the line that connects them crosses the Milky Way at an angle. This is due to the ecliptic being tilted at about 60 degrees to the galactic plane on which our entire Milky Way lies. If the Milky Way could somehow be shrunk down to lie flat on a table, our Solar System would be jutting out like a pin stuck in it at an odd angle.